\documentclass[12pt,a4page]{article}
\usepackage{epsfig}
\begin{document}
\fontsize{11}{11}\selectfont 
\title{Phenomenological Parameters of the Prototype Eclipsing Binaries Algol, $\beta$ Lyrae and W UMa}
\author{\textsl{M.\,G.~Tkachenko$^{1}$}, I.\,L.~Andronov$^{1}$, L.\,L.~Chinarova$^2$}
\date{\vspace*{-6ex}}
\maketitle
\begin{center} {\small $^{1}$Department "Mathematics, Physics and Astronomy", Odessa National Maritime University, Mechnikova Str., 34, 65029, Odessa, Ukraine\\
{\tt masha.vodn@yandex.ua, tt\_ari@ukr.net}}\\
{\small $^{2}$ Astronomical Observatory, Odessa National University, Marazlievskaia Str., 1V, 65014, Odessa, Ukraine\\
{\tt llchinarova@gmail.com}}\\
\end{center}

\begin{abstract}
The phenomenological parameters of eclipsing binary stars, which are the prototypes of the EA, EB and EW systems are determined using the expert complex of computer programs, which realizes the NAV ("New Algol Variable") algorithm (Andronov 2010, 2012) and its possible modifications are discussed, as well as constrains for estimates of some physical parameters of the systems in a case of photometric observations only, such as the degree of eclipse, ratio of the mean surface brightnesses of the components. The half-duration of the eclipse is 0.0617(7), 0.1092(18) and 0.1015(7) for Algol, $\beta$ Lyrae and W UMa, respectively. The brightness ratio is 6.8$\pm$1.0, 4.9$\pm$1.0 and 1.15$\pm$0.13. These results show that the eclipses have distinct begin and end not only in EA (as generally assumed), but also in EB and EW - type systems as well. The algorithm may be applied to classification and study of the newly discovered (or poorly studied) eclipsing variables based on own observations or that obtained using photometric surveys.
\\[1ex]
\end{abstract}

\section*{\sc introduction}

Modern photometric surveys have lead to discovery of hundred thousands of new variable stars, which are subject of robust classification and determination of the main phenomenological parameters, which are necessary for a registration in the General Catalogue of Variable Stars" \cite{samus} or in the AAVSO "Variable Stars Index" \cite{aavso} - the type, brightness at maxima and minima, period, initial epoch, duration of eclipse in per cent (for eclipsing binaries), or asymmetry $M-m$ for pulsating variables. In the "remarks" section for the eclipsing variables, it is needed to list an amplitude of the secondary minimum, duration of the total eclipse.

One may note previous extensive studies in Ukraine of Algols \cite{shulberg}, $\beta$ Lyr \cite{skulsky} \cite{naz}, W UMa - type stars \cite{karet} (for more details, see classical monographs \cite{tsesevich}, \cite{kopal1959}). More recent reviews and monographs are published in \cite{ruc1993}, \cite{kallrath2009}. These studies are carried out in a frame of the international campaign "Inter-Longitude Astronomy" (\cite{andr2003a},\cite{andr2010a}) and national projects "Ukrainian Virtual Observatory" and "Astroinformatics" (\cite{vav2011},\cite{vav2012}).   

One of the most common methods for the analysis of periodic signals is a trigonometric polynomial approximation (sometimes referred to as a truncated Fourier series). Statistically optimal degree of trigonometric polynomial is often small for "almost sinusoidal" signals (e.g. pulsating variables \cite{kuda1996}, \cite{am2006}) but increases significantly if the signal has intervals of rapid change \cite{a1994}, \cite{andr2003}. For Algols, which are characterized by relatively narrow eclipses, this is the case, which can be observed in the phase light curves of eclipsing binary systems, especially of the Algol type. The modelling of the light curves of eclipsing binaries using the Fourier coefficients was discussed by \cite{ruc1998} (mainly for EW - type stars), \cite{kopal1978} (mainly for Algol - type stars) and references therein.

Although the Fourier coefficients may be determined with an excellent accuracy for the theoretical light curves, the accuracy of the coefficients for the really observed light curves is much worse either due to the observational errors, or due to inhomogeneity of coverage of the light curve by observations. The usual simplified equations for the Fourier coefficients should be replaced by complete equations for the least squares (LSQ) \cite{a1994}, \cite{andr2003}, \cite{mikulas2007}.

Another approach is to use special shapes (also called "patterns" or "profiles") of the eclipses. The simplified model of spherical components with a constant brightness distribution is widely used (\cite{shulberg}, \cite{malkov2007}, \cite{tk2013} for a preliminary determination of the parameters. Also there is an approach to compute "physical" models based on the approach proposed by Wilson and Devinney \cite{wd71} realized by various authors (\cite{wd94}, \cite{zola1}, \cite{zola2}, \cite{phoebe}, \cite{bm}), also for more complicated cases with accretion disks (\cite{cher1993}).  

However, the real physical modelling assumes additional spectroscopic observations to determine orbital velocities of the components, the mass ratio, and temperature(s). 
This is available for $\leq 1\%$ of the known objects, so for the rest 
the phenomenological modelling is the only source of information.

Fixing values of some of these unknown parameters (temperature at the pole of one component and the mass ratio), it is possible to compute the rest of them. However, the (statistically) same quality of the approximation may be obtained for a region in the parameter space, rather than at some "statistically optimal" point. Such computations need much longer computational time, the error estimates need even much more resources, thus the "phenomenological" approximations using simpler functions remain much more effective in a sense "less computer resources to get a same accuracy".

\section*{\sc Phenomenological Models}

Generally, the phenomenological parameters may be determined using two classes of the approximations: the "local" and "global" ones. The first class is based on the "local" approximations of the extrema, historically starting from "hand-written" approximation of the points on a millimeter paper (see e.g. \cite{tsesevich1980} for a review). Later on, the intervals near the extrema were approximated using algebraic polynomials of the user-defined degree (often an ordinary parabola \cite{tsesevich}, \cite{tsesevich1980}, \cite{andr2000}, \cite{papa2014}) or using the statistically optimal degree (\cite{b2003}, \cite{andrandr2015}). Other advanced methods were reviewed by \cite{andr2005}, \cite{mikulas2015}, \cite{andr2016b}. 

The trigonometric polynomial (also called the "restricted Fourier series") is expressed as
\begin{eqnarray}
 x_c(\phi)&=& C_1+\sum_{j=1}^s(C_{2j}\cos(2\pi j\phi)+C_{2j+1}\sin(2\pi j\phi))\nonumber\\
&=&C_1+\sum_{j=1}^s R_j\cos(2\pi j(\phi-\phi_{0j})),
\label{eq1}
\end{eqnarray}
where $\phi=\psi-{\rm int}(\psi),$ ${\rm int}(\psi)$ is an integer part of $\psi,$ $\psi=(t-T_0)/P$ is phase, $t$ is time, $T_0$ is the initial epoch, and $P$ is the period (cf. \cite{a1994}, \cite{andr2003}. If needed, the period may be improved using differential corrections.

The "symmetrical" trigonometric polynomial contains only terms with cosines, the coefficients $C_{2j+1}$ of the terms with sines are suggested to be zero (e.g. if their deviations from zero are not statistically significant, one may fix these values to zero), as realized in the program MCV \cite{ab04}. If they are statistically significant, the maxima are unequal, what is called the O'Connell effect (\cite{oc}).

For the light curves of eclipsing variables, it may be effective to split the trigonometric polynomial into two parts:
\begin{eqnarray}
x_m(\phi)=\frac{x_c(\phi)+x_c(\phi+0.5)}{2}&=& C_1+\sum_{k=1}^{s/2}(C_{4j}\cos(4\pi k\phi)+C_{4j+1}\sin(4\pi k\phi)),\\
x_d(\phi)=\frac{x_c(\phi)-x_c(\phi+0.5)}{2}&=& \sum_{k=1}^{s/2}(C_{4j-2}\cos(2\pi (2k-1)\phi)+C_{4j-1}\sin(2\pi(2k-1)\phi))
\nonumber
\label{eq2}
\end{eqnarray}
\cite{tk2016}. Obviously, their sum corresponds to $x_c(\phi)$ and the difference -- to $x_c(\phi+0.5),$ respectively. If the O'Connell effect is absent, the "mean" function $x_m(\phi)$ describes the out-of-eclipse part of the light curve, and twice - the minima "of equal depth and shape". The difference between the minima are described by the "deviation" function $x_d(\phi),$ which, in a case of EW-type stars is very close to zero at all phases. For statistically best approximation, one has to use differential corrections to determine not only the period, but also the initial epoch.

For the Algol-type stars, the statistically significant degree of the trigonometrical polynomial $s,$ computed using the Fischer's criterion, may reach 21, leading to a huge number of parameters $m=2s+2=44$ (e.g. \cite{andr2016b}. Another star was best characterized by even larger value $s=50$ (\cite{andr2016a}), showing a strong Gibbs phenomemon.

Andronov (\cite{andr2010},\cite{andr2012}) proposed a non-polynomial spline approximation
\begin{eqnarray}
x_c(\phi)&=& C_1+C_2\cos(2\pi\phi)+C_3\sin(2\pi\phi)+\nonumber\\
&& +C_4\cos(4\pi\phi)+C_5\sin(4\pi\phi)+\\
&& +C_6H(\phi, C_8,C_9)+C_7H(\phi+0.5, C_1, C_{10}).\nonumber   \label{eq11}
\end{eqnarray}
Additionally, we define a particular sum $x_{c5}(\phi),$ which contains only 5 terms without taking into account the contributions (to the stellar magnitude) of the primary (6-th term) and secondary (7-th term) eclipses.

Here, for suitability of computations, the phase is redefined to be in the interval $[-0.25,+0.75)$ instead of the usual definition in the main interval $[0,1).$ Moreover, the phase may be corrected as $\phi=\tilde{\phi}-C_{11}-C_{12}(t-T_0)$ to take into account possible corrections for the initial epoch and the period (see \cite{tk2013} for more details). Here $\tilde{\phi}$ is the phase corresponding to the initial values of $T_0$ and $P.$

The basic pattern (shape) is 
\begin{equation}
 H(\zeta,C_8,\beta)=\left\{
\begin{array}{ll}
V(z)=(1-|z|^\beta)^{3/2},  & {\rm if} |z|<1,\\
 0,  & {\rm if} |z| \geq 1
\end{array} 
 \right.
 \label{eq12}
\end{equation}
where a dimensionless parameter $z=\phi/C_8$.

So, the parameters have following meanings:

 \begin{tabular}{ll}
  $C_{1}-$ & the mean of the $x_{c5}(\phi)$ over a complete phase interval;
\\ $C_{2}-$ & the semi-amplitude of the reflection effect;
\\ $C_{3}, C_{5}-$ &the semi-amplitudes of the sine terms, which describe the O'Connell effect;
\\ $C_{4}-$ & the semi-amplitude of the effect of ellipticity;
\\ $C_{6}-$ & the amplitude of the primary minimum;
\\ $C_{7}-$ & the amplitude of the secondary minimum;
\\ $C_{8}-$ & the eclipse half - duration (the phase of the end of eclipse);
\\ $C_{9}-$  & the parameter describing the shape of the primary minimum;
\\ $C_{10}-$ & the parameter describing the shape of the secondary minimum;
\\ $C_{11}-$ & the phase correction;
\\ $C_{12}-$ & the frequency correction. \\
  \end{tabular}

The first 7 parameters may be determined using the linear least squares method, but other 7 "non-linear" parameters $C_8-C_{12}$ may be determined using the differential corrections after a "brute force" minimization of the test (target) function
\begin{equation}
\Phi_m=\sum_{k=1}^{n}w_k\cdot(x_k-x_c(\phi_k))^2.\nonumber
\label{eq3}
\end{equation}
The "unit weight" error $\sigma_{0m}=\sqrt{\Phi_m/(n-m)},$ and the r.m.s. accuracy of the approximation at the phases of observations is $\sigma[x_c]=\sigma_{0m}\sqrt{m/n}.$ In the current version of the program, it is possible to fix some parameters, not permitting corrections to them. The properties of the test function were discussed in \cite{tk2016}.

In the previous versions, the Monte-Carlo method was used for the minimization of the test function in the parameter space, but it needs much more computational time as compared to the combination of the "brute force"$+$"differential corrections" (\cite{andr2013}).

The light curves of the prototype stars Algol, $\beta$ Lyr and W UMa are shown in Figures 1,2,3, respectively, and the best fit parameters are listed in Table 1. We have used the published photoelectric observations for these stars, the references are shown in the figure captions. For two stars, the observations were obtained in two filters (B and V for $\beta$ Lyr and blue ("B") and yellow ("Y") for W UMa), so we used our program to these data files separately. 

The only parameter, which is expected to be the same for all filters, is $C_8.$ For the analysed data, the difference between the estimates obtained for different filters, is not statistically significant for the examined stars. Also one may note a significantly narrower eclipse in Algol $(C_8=0.062)$ as compared to $\beta$ Lyr $(C_8=0.105)$ and W UMa $(C_8=0.101).$ As the eclipse duration is dependent on the sum of relative radii of the components and the inclination, the EB and EW - type stars may have small durations, but the EA - type stars typically have $C_8\leq 0.08.$

 For the "fine tuning", one has to minimize the weighted sum of the test functions, which depend on this joint parameter (the rest are computed to minimize the test function after fixing the trial value of this parameter). Such an approach we previously used in \cite{a2015}.

In an addition to the best fit parameters, we use additional combinations of them, which are related to the physical parameters of the stars. Among them are $d_1=1-10^{-C_6}$ and  $d_2=1-10^{-C_7} -$ the ratio of the deficit of flux at the eclipse (primary and secondary, respectively) to the theoretically expected value assuming that the obscuration is absent. This may be significantly different from typically used value of the amplitude defined as $\Delta=x_c(\phi_{min})-x_c(\phi_{max}),$ because we take into accounts the reflection, ellipticity and the O'Connell effects, thus $C_6=x_c(0)-x_{c5}(0),$  $C_7=x_c(0.5)-x_{c5}(0.5).$ The differences in the estimates of the eclipse depths using these methods may reach dozens per cent for the EW-type stars, and, for the elliptic-type stars (no eclipses), our method will indicate $C_6\approx 0$ and  $C_7\approx 0$ within the error estimates. The classical amplitude will remain detectable  $\Delta\approx 0.\!^{\rm m}2.$

There are two parameters, which are related to $d_1,$ $d_2,$ namely, their sum $Y=d_1+d_2$ and $\zeta=d_1/d_2$ (\cite{a2015}, \cite{andr2016b}). The first one characterizes the presence of eclipse ($Y=0$ if no eclipse, and $Y=1$ if both eclipses are total). The brightness ratio $\zeta$ indicates the relative temperatures of the components. Using these parameters for both filters, and the statistical relationships Mass-Luminosity-Radius, we \cite{a2015} estimated the physical parameters of the newly discovered system 2MASS J$18024395+4003309$ = VSX J$180243.9+400331.$

Six other newly discovered stars show a presence of the contribution of eclipses not only for the EA - type, but also for the EW \cite{tk2015}.

For the stars Algol and $\beta$ Lyr, we have used the phases computed by the authors of the corresponding papers, thus no period correction was made, thus the parameter $C_{12}$ was set to zero. For W UMa, the corrections are small, but may be determined.

A separate remark should be done on the fixed values of $C_9$ or $C_{10}.$ From the model of spherical stars with uniform brightness distribution, one may expect the minimal limit of $\beta$ in Eq.(\ref{eq12}) of 1.5, which corresponds to a coincidence of both inner contacts of equal stars at an inclination $i=90^\circ.$ In other cases, this value is expected to be larger. If the value of the parameter after any iteration formally goes outside the user-defined limits, this value is set to the corresponding nearest limit (minimal or maximal), and the parameter becomes fixed at the limit, unless next iteration will move it inside the "permitted" interval. 

However, for the examined stars, the minima are sharp enough to make this parameter be equal to the minimal limit either for Algol, or to W UMa. For $\beta$ Lyr, the parameter $C_9$ is the same within error estimates for both filters, whereas $C_{10},$ which corresponds to the secondary minimum, has a very large error estimate. Analysing the Fig 2, one may suggest that this is due to a shallow minimum and a relatively small number of points. Despite this parameter is unsure, the corresponding depth of the minima is characterized by a better accuracy estimate.

\section*{\sc Improved Approximations}

As the shape parameters $\beta$ are often equal to the minimal limit 1.5 for the examined stars, as well as for the previously studied stars, we have applied approximations with additional parameters. The dozens of modified functions were tested by \cite{andr2016a}, \cite{andr2016c}. In Figure 4, there are shown the part of the approximation near the primary minimum with the main set of parameters listed in Table 1, as well as modifications. One of the modifications is for the value $C_9=1,$ which corresponds to a triangular shape at the center of the eclipse. This triangle makes an apparent enlarging of the eclipse depth, however, which is not statistically different from the previous approximation. Another modification was found to be the best by \cite{andr2016c}. It redefines $z$ in Eq.(\ref{eq12}) as $z=y+C_{13}y(1-y),$ where $y=|\phi/C_8|$ for the primary minimum and  $z=y+C_{14}y(1-y),$ where $y=|(\phi-0.5)/C_8|$ for the secondary minimum. The obvious restrictions are $0\leq y\leq 1,$ $-1\leq C_{13}\leq 1.$ The depth of the minimum is the same within error estimates, and there is a good approximation of both the ascending and descending branches.

\section*{\sc Conclusion}

We have applied the algorithm NAV ("New Algol Variable") to three prototype stars and obtained phenomenological parameters, which may be used for comparison with that for other stars. The method is effective not only for the EA-type systems, for which it was mainly elaborated, but also for EB and even EW -type stars. The revised definition of the depth of the minimum with taking into account effects of proximity (reflection, ellipticity and O'Connell) leads to more precise determination of the ratio of the mean brightnesses of the eclipsed parts of the components, which is useful for the physical modelling. The current algorithm may be used for determination of phenomenological parameters of numerous new variables discovered from the ground-based and space surveys.


\begin{table*}
 \caption{Parameters of the approximation (\ref{eq1}).}\label{tab1}
 \vspace*{1ex}
 {\footnotesize
 \begin{tabular}{cccccc}
  \hline 
  Par. & Algol & $\beta$ Lyr (B) & $\beta$ Lyr (V)  & W UMa (B) & W UMa (Y) \\
  \hline
 $C_1$ &  0.6597 $\pm$ 0.0031 & 3.5498 $\pm$ 0.0048 & 3.5427 $\pm$ 0.0058 & -1.1923 $\pm$ 0.0015 & -1.0161 $\pm$ 0.0016 \\
 $C_2$ &  0.0466 $\pm$ 0.0043 & 0.0024 $\pm$ 0.0058 & -0.0051 $\pm$ 0.0069 & 0.0084 $\pm$ 0.0016 & 0.0097 $\pm$ 0.0016 \\
 $C_3$ &  0.0045 $\pm$ 0.0044 & 0.0067 $\pm$ 0.0034 & 0.0100 $\pm$ 0.0038 & -0.0252 $\pm$ 0.0012 & -0.0205 $\pm$ 0.0011 \\
 $C_4$ &  0.0195 $\pm$ 0.0052 & 0.1565 $\pm$ 0.0072 & 0.1420 $\pm$ 0.0087 & 0.1583 $\pm$ 0.0025 & 0.1412 $\pm$ 0.0026 \\
 $C_5$ &  0.0032 $\pm$ 0.0038 & -0.0204 $\pm$ 0.0047 & -0.0195 $\pm$ 0.0051 & 0.0155 $\pm$ 0.0017 & 0.0116 $\pm$ 0.0016 \\
 $C_6$ &  0.9786 $\pm$ 0.0111 & 0.6522 $\pm$ 0.0159 & 0.6259 $\pm$ 0.0900 & 0.4132 $\pm$ 0.0054 & 0.4068 $\pm$ 0.0053 \\
 $C_7$ &  0.0996 $\pm$ 0.0152 & 0.1067 $\pm$ 0.0177 & 0.1016 $\pm$ 0.0226 & 0.3440 $\pm$ 0.0048 & 0.3444 $\pm$ 0.0049 \\  
 $C_8$ &  0.0617 $\pm$ 0.0007 & 0.1093 $\pm$ 0.0018 & 0.1092 $\pm$ 0.0022 & 0.0999 $\pm$ 0.0009 & 0.1029 $\pm$ 0.0010 \\
 $C_9$ &  1.5$^*$ & 2.0370 $\pm$ 0.0862 & 2.0216 $\pm$ 0.0981 & 1.7733 $\pm$ 0.0442 & 1.6835 $\pm$ 0.0404 \\ 
 $C_{10}$ &  1.5$^*$ & 1.5$^*$  & 2.8990 $\pm$ 1.8908 & 1.5$^*$ & 1.5$^*$  \\ 
 $C_{11}$ &  0.0009 $\pm$ 0.0003 & 0.0030 $\pm$ 0.0007 & 0.0032 $\pm$ 0.0007 & 0.0005 $\pm$ 0.0003 & 0.0010 $\pm$ 0.0003 \\
 $C_{12}$ &  0$^*$ & 0$^*$  & 0$^*$  & $(-202 \pm 85)\cdot 10^{-7}$ & $(-386 \pm 85) \cdot 10^{-7}$ \\
 $d_1$ &  0.5940 $\pm$ 0.0042 & 0.4516 $\pm$ 0.0080 & 0.4381 $\pm$ 0.0098 & 0.3165 $\pm$ 0.0034 & 0.3125 $\pm$ 0.0034 \\
 $d_2$ &  0.0876 $\pm$ 0.0128 & 0.0936 $\pm$ 0.0148 & 0.0893 $\pm$ 0.0190 & 0.2716 $\pm$ 0.0032 & 0.2718 $\pm$ 0.0033 \\
 $Y$ &  0.6816 $\pm$ 0.0139 & 0.5452 $\pm$ 0.0194 & 0.5274 $\pm$ 0.0241 & 0.5881 $\pm$ 0.0057 & 0.5843 $\pm$ 0.0057 \\
 $\gamma$ &  6.7773 $\pm$ 0.9859 & 4.8233 $\pm$ 0.7294 & 4.9060 $\pm$ 1.0119 & 1.1656 $\pm$ 0.0140 & 1.1495 $\pm$ 0.0134 \\
 $\sigma_0$ &  0.0232 & 0.0294 & 0.0283 & 0.0173 & 0.0167 \\
  \hline
  \end{tabular}
}
The asterics mark fixed parameters, which were not optimized.
\end{table*}

\begin{figure}[!h]
\centering
\epsfig{file = 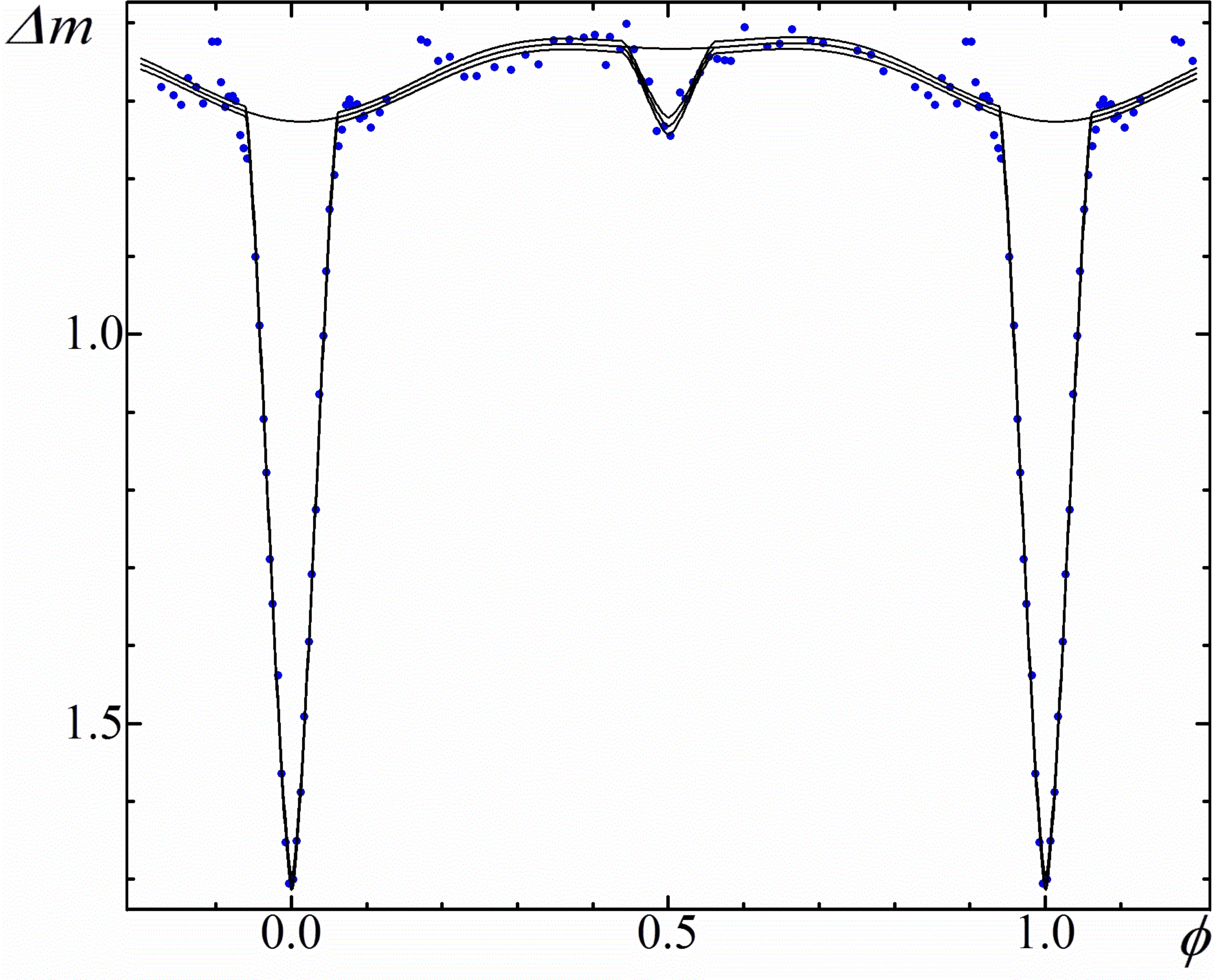,width = 0.5\linewidth}
\caption{Phase light curve of Algol ($\beta$ Per). The points are observations by \cite{cris1966}, the lines are the NAV approximation with $\pm 1\sigma$ "error corridor".}
\label{fig1}
\end{figure}

\begin{figure}[!h]
\centering
\epsfig{file = 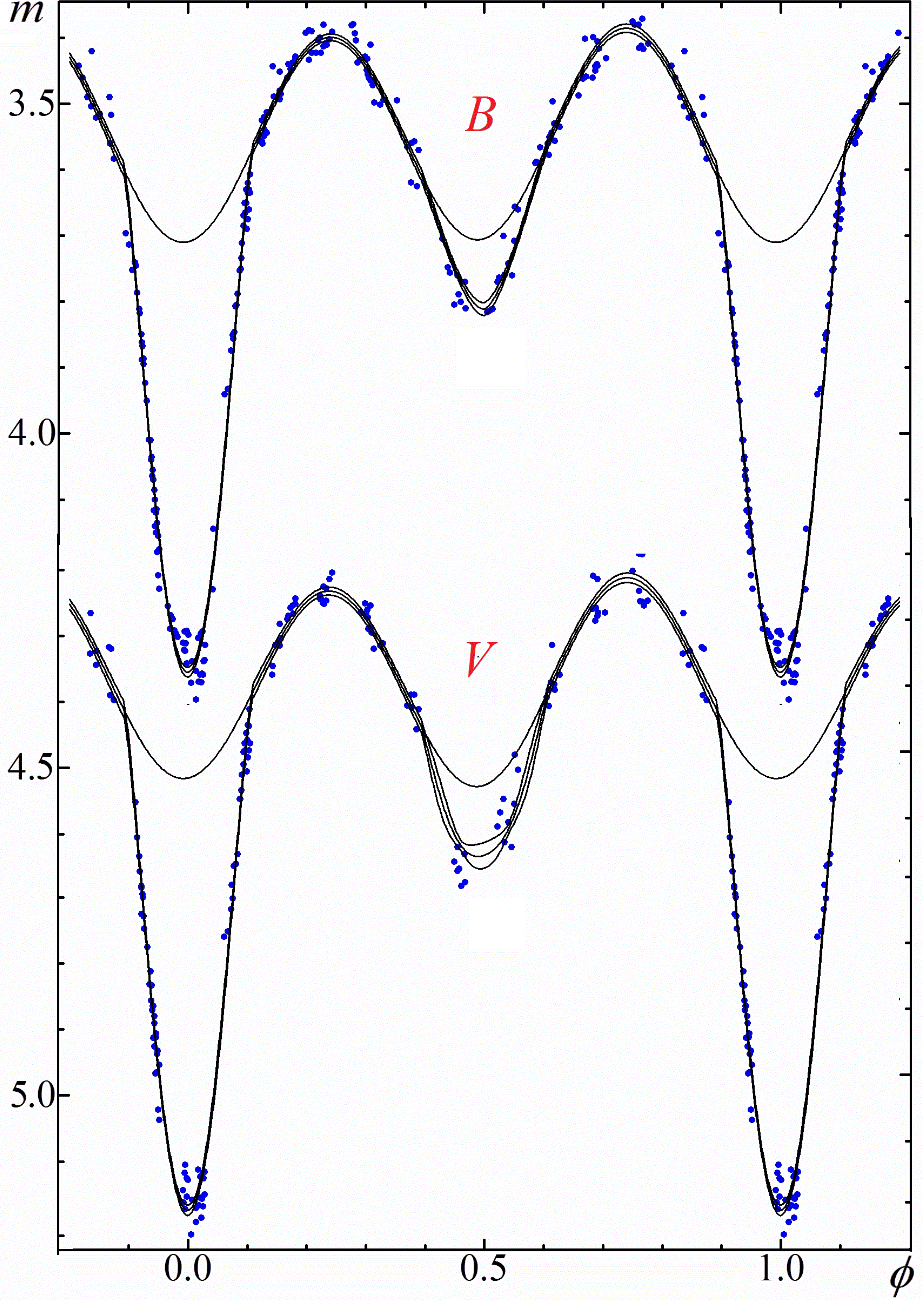,width = 0.5\linewidth}
\caption{Phase light curve of $\beta$ Lyr in BV filters. The points are observations by \cite{lars1969}, the lines are the NAV approximation with $\pm 1\sigma$ "error corridor".}\label{fig2}
\end{figure}

\begin{figure}[!h]
\centering
\epsfig{file = 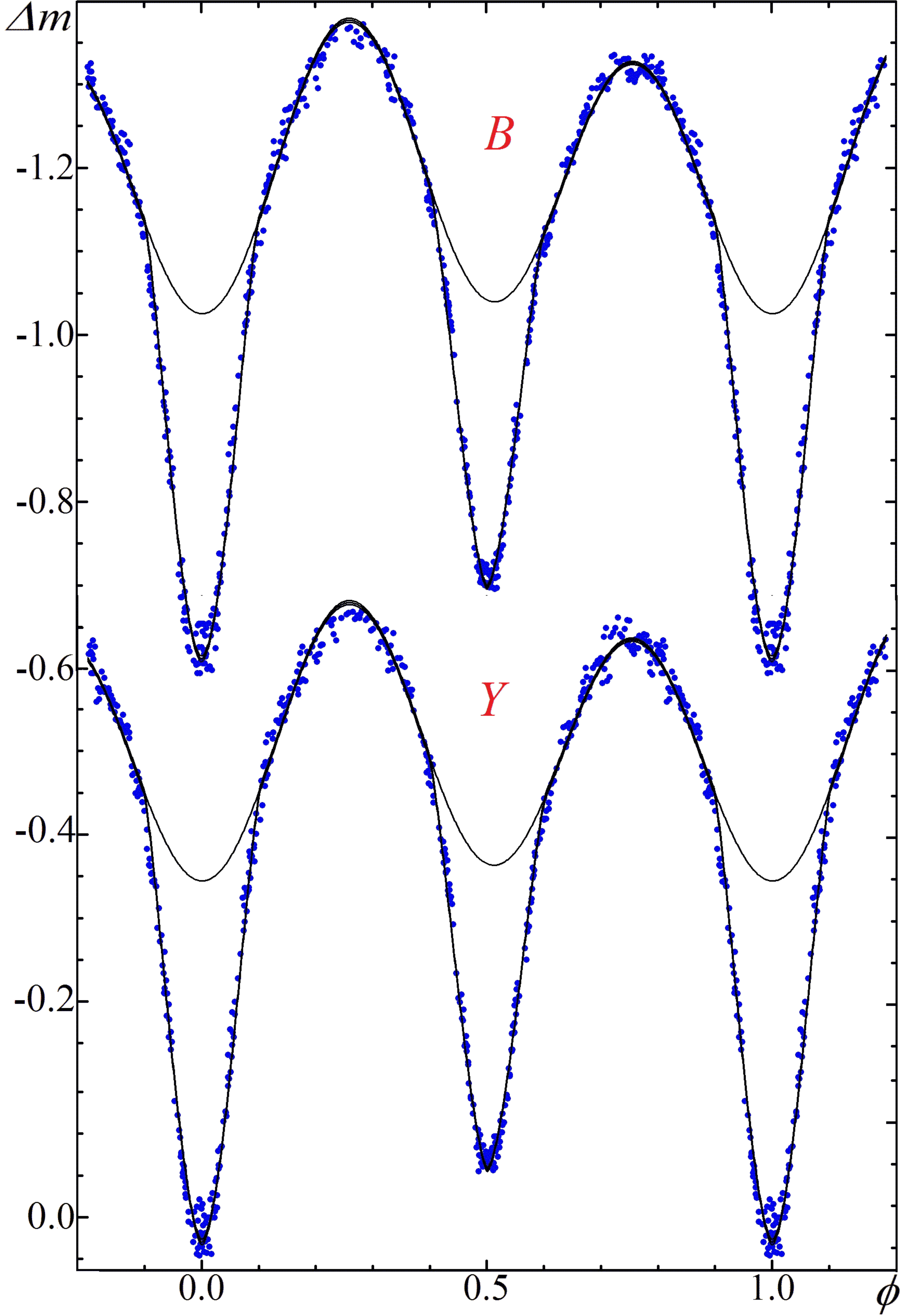,width = 0.5\linewidth}
\caption{Phase light curve of W UMa in blue (B) and yellow (Y) filters. The points are observations by \cite{bin1966}, the lines are the NAV approximation with $\pm 1\sigma$ "error corridor".}\label{fig3}
\end{figure}

\begin{figure}[!h]
\centering
\epsfig{file = 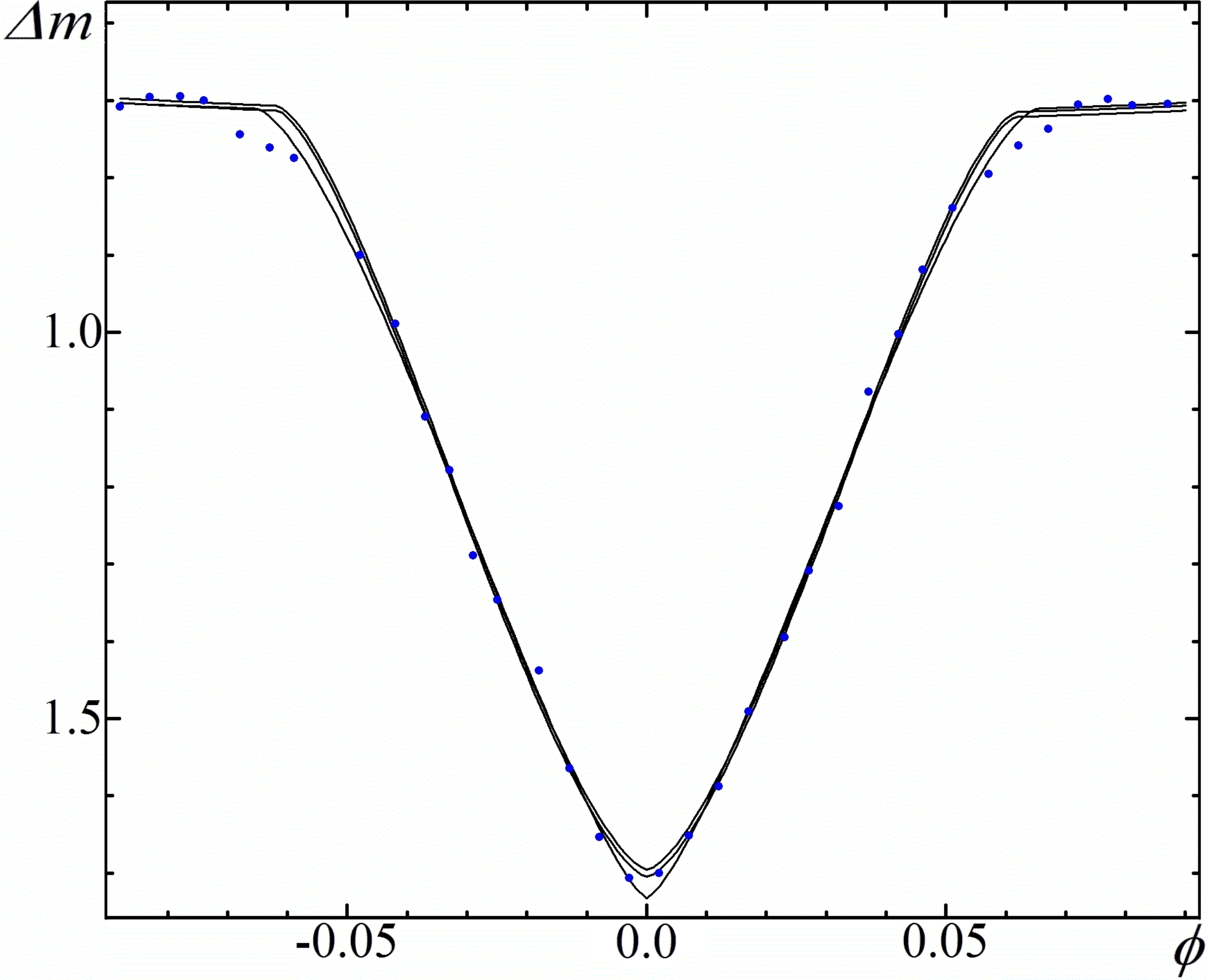,width = 0.5\linewidth}
\caption{Phase light curve of Algol ($\beta$ Per) near the primary minimum. The points are observations by \cite{cris1966}, the lines are the NAV approximations, which correspond to 3 different sets of the parameters: $C_9=1.5,$ $C_{13}=0.8,$ (up at the mid-eclipse),
$C_9=1.5,$ $C_{13}=0,$ (middle),
 $C_9=1,$ $C_{13}=C_{14}=0$ (bottom).}\label{fig4}
\end{figure}

\end{document}